\documentclass[11pt,a4paper]{article}

\pdfoutput=1
\usepackage[T1]{fontenc}
\usepackage[utf8]{inputenc}
\usepackage{lmodern}
\usepackage[a4paper,margin=1in]{geometry}
\usepackage{amsmath,amssymb,mathtools}
\usepackage{bm}
\usepackage{physics}
\usepackage{booktabs}
\usepackage{microtype}
\usepackage{graphicx}
\usepackage[numbers,sort&compress]{natbib}
\usepackage{hyperref}
\usepackage{enumitem}

\hypersetup{
  colorlinks=true,
  linkcolor=blue,
  citecolor=blue,
  urlcolor=blue,
  pdftitle={A Braak-Inspired Spectral Determinant Approach to a Dirac Bound-State Problem},
  pdfauthor={S. Zarrinkamar}
}

\allowdisplaybreaks[3]

\begin{document}

\begin{flushright}
\small PREPARED FOR SUBMISSION TO JHEP
\end{flushright}

\title{A Braak-Inspired Spectral Determinant Approach to a Dirac Bound-State Problem}

\author{S.~Zarrinkamar}
\date{}

\maketitle

\begin{center}
\small
Department of Basic Sciences, Ga. C., Islamic Azad University, Garmsar, Iran\par
E-mail: \href{mailto:saber.zarrinkamar@iau.ac.ir}{saber.zarrinkamar@iau.ac.ir};\quad \href{mailto:zarrinkamar.s@gmail.com}{zarrinkamar.s@gmail.com}\par
\href{https://orcid.org/0000-0001-9128-4624}{ORCID: 0000-0001-9128-4624}
\end{center}

\begin{abstract}
We develop a parity-resolved spectral-connection construction for a one-dimensional Dirac particle in an attractive soft-core Coulomb potential. The square-root dependence of the potential produces algebraic branch points in the coordinate plane. We uniformize these branch points by introducing \(t=\sqrt{x^2+\beta^2}\), while retaining the full-line \(Z_2\) symmetry generated by \(\Pi=\sigma_z P_x\). In each parity sector the resulting problem is formulated as a global connection problem between a solution selected at the physical origin and the decaying solution at infinity. This defines parity-resolved spectral connection functions whose zeros determine bound-state energies. We further construct a quasi-exact Bethe-ansatz sector directly from the same Dirac equations and use its polynomial solutions as independent benchmarks of the spectral condition. The comparison clarifies the distinction between local quasi-exact solvability and the global spectral determinant. The construction is inspired by the role of discrete symmetry and global analyticity in the quantum Rabi model, without identifying the present spectral function with Braak's \(G\)-function.
\end{abstract}


\section{Introduction}

Exact analytical solutions occupy a special place in quantum mechanics. They provide direct access to spectra, eigenfunctions, symmetry structures, and limiting regimes that are often difficult to isolate by purely numerical methods. At the same time, genuinely exactly solvable quantum systems are relatively rare. Even within supersymmetric quantum mechanics (SUSYQM), where factorization, intertwining relations, shape invariance, and algebraic constructions considerably enlarge the class of analytically tractable Hamiltonians, exact solvability is restricted to special families of potentials and symmetry structures \cite{CooperKhareSukhatme1995,CooperKhareSukhatme2001,Junker1996,Bagchi2001}. The algebraic and factorization viewpoints developed in this context have also provided useful connections between nonrelativistic and relativistic wave equations and between apparently different spectral problems. \\
The development of SUSYQM has consequently generated a broad literature on exact solutions and on the systematic generation of solvable partner Hamiltonians. The standard treatments by Cooper, Khare, and Sukhatme \cite{CooperKhareSukhatme1995,CooperKhareSukhatme2001}, Junker \cite{Junker1996}, and Bagchi \cite{Bagchi2001} emphasize factorization, shape invariance, intertwining operators, and algebraic methods. Related developments include the relativistic realization of supersymmetric structures \cite{HughesKosteleckyNieto1986,ZnojilKG2004}, the application of supersymmetric ideas to atomic spectra \cite{KosteleckyNieto1984}, and factorization methods for relativistic and nonrelativistic wave equations \cite{Dong2007,Dong2011}. Exact Dirac-type solutions obtained through factorization have also been studied in magnetic backgrounds and related systems \cite{KuruNegroNieto2009}. These results illustrate an important general lesson: exact solvability is usually associated with a special algebraic, differential, or symmetry structure rather than with a generic potential.\\
This observation motivates the present problem. The Hamiltonian considered here contains a physically motivated soft-core Coulomb interaction and does not belong, in any obvious way, to the standard shape-invariant families. The presence of the square root \(\sqrt{x^2+\beta^2}\) introduces algebraic branch points and makes a direct reduction to elementary or hypergeometric functions nontrivial. Soft-core Coulomb interactions of the form used here have been studied extensively, including relativistic treatments based on spectral bounds and quasi-exact Bethe-ansatz constructions \cite{HallZorin2012,AgboolaZhang2012}. In the relativistic literature, the closest formulations generally involve higher-dimensional central potentials or simultaneous scalar and vector couplings. One-dimensional Dirac problems with even attractive potentials have also been discussed in general terms \cite{HallZorinNodal2014}. The present work instead considers the full line with a pure time-like vector interaction $V(x)I$, corresponding to an electrostatic coupling and uses its exact spinor-parity symmetry explicitly. Such features are precisely where the distinction between exact solvability, quasi-exact solvability, and a global spectral construction becomes useful. Supersymmetric methods have also been extended beyond conventional Hermitian settings and to systems with nontrivial analytic structure, including PT-symmetric and pseudo-Hermitian formulations \cite{Znojil2001,Znojil2004,FringKorff2004,FringFrith2019}. These developments are relevant here not because the present Hamiltonian is PT symmetric or supersymmetric, but because they demonstrate how factorization and analytic continuation can expose otherwise hidden spectral structures.\\
These developments were pursued extensively by researchers at the University of Valladolid. Negro, Nieto, and collaborators studied factorization and confluent-hypergeometric structures of solvable quantum potentials, including Coulomb and Morse systems, and developed refined factorizations of solvable potentials \cite{NegroNietoRosas2000,NegroNietoRosas2000b}. Related work on second-order Darboux transformations and periodic systems clarified how supersymmetric transformations can preserve or reorganize spectral structures \cite{FernandezNegroNieto2000,NegroNietoRosas2004}. In the relativistic setting, Nieto, Pecheritsin, and Samsonov developed intertwining constructions for the one-dimensional Dirac equation, including polynomial factorizations and quadratic supersymmetry \cite{NietoPecheritsinSamsonov2003}, while Kuru, Negro, and Nieto applied factorization methods directly to Dirac-Weyl systems \cite{KuruNegroNieto2009}. These works are particularly relevant to the present study because they illustrate how SUSYQM and intertwining methods can remain useful even when the underlying wave equation is relativistic or when the potential is not among the simplest textbook shape-invariant families.\\
A related and particularly important class is provided by quasi-exactly solvable (QES) quantum systems. In a QES problem, only a finite portion of the spectrum and the associated eigenfunctions can be obtained algebraically for special parameter values. The modern theory of QES models was developed through the work of Turbiner and collaborators, who connected finite-dimensional invariant spaces with Lie-algebraic structures such as \(sl(2)\) \cite{Turbiner1987,Turbiner1988}. The systematic treatment by Ushveridze emphasized the broad role of QES models, including their Bethe-ansatz, Lie-algebraic, and analytic formulations \cite{Ushveridze1994}. Necessary and sufficient conditions for one-dimensional QES Schr\"odinger operators were subsequently studied by González-López, Kamran, and Olver \cite{GonzalezLopezKamranOlver1993,GonzalezLopezKamranOlver1994}. The relation between QES systems and orthogonal polynomials, together with the emergence of finite-dimensional spectral problems, was further developed by Bender and Dunne \cite{BenderDunne1996}, while semiclassical aspects of the boundary between solvable and nonsolvable parts of the spectrum were considered by Bender, Dunne, and Moshe \cite{BenderDunneMoshe1996}. These approaches make clear that QES solvability is not simply a weaker version of exact solvability: it is a distinct algebraic structure in which a finite invariant sector survives inside a generally non-solvable spectral problem. Functional Bethe-ansatz methods provide an especially direct realization of this finite-sector algebraization. Zhang developed polynomial Bethe-ansatz constructions for second-order differential equations and applied them to the quantum Rabi model and several of its generalizations, obtaining algebraic conditions for the roots and the associated exceptional energies \cite{ZhangPolynomial2012,ZhangRabi2013,AgboolaZhang2013,AgboolaZhang2012}. This line of work is particularly relevant here because it separates a finite algebraic sector from the full spectral problem. The distinction is especially relevant for relativistic systems. Quasi-exact solutions can provide exact algebraic benchmarks without implying that the entire spectrum is accessible in closed form. They can therefore be used to test more general spectral constructions. In the present work we exploit this distinction: the Bethe-ansatz solutions are not taken to define the spectrum, but rather provide independent analytic points at which the global spectral connection function can be tested. \\
Another major route to analytical quantum mechanics is provided by special-function reductions. The hypergeometric and confluent-hypergeometric equations cover many familiar exactly solvable problems, but more complicated potentials frequently lead to Heun-type equations. The general Heun equation is the canonical second-order Fuchsian equation with four regular singular points, while its confluent, doubly confluent, biconfluent, and triconfluent reductions arise when singularities coalesce. The mathematical theory and physical applications of these equations are collected in the classic volume edited by Ronveaux \cite{Ronveaux1995} and in the monograph by Slavyanov and Lay \cite{SlavyanovLay2000}. Applications to quantum mechanics and mathematical physics have continued to expand, as reviewed, for example, by Horta\c{c}su \cite{Hortacsu2018}. The appearance of Heun equations is particularly natural when several finite singularities coexist with nontrivial asymptotic behavior, precisely the type of analytic structure encountered after reductions of many relativistic and nonrelativistic wave equations. \\
Recent work by Ishkhanyan and collaborators has further demonstrated the breadth of this connection. Exact Schr\"odinger potentials reducible to the general Heun equation have been systematically classified \cite{Ishkhanyan2016}, while special classes of Heun solutions have been expressed through generalized hypergeometric functions under appropriate restrictions on the characteristic exponents and accessory parameters \cite{Ishkhanyan2018}. The inverse-square-root potential provides a particularly instructive example: its Schr\"odinger equation admits a representation in terms of a triconfluent Heun function and also exhibits quasi-polynomial solutions \cite{Ishkhanyan2015}. Such results illustrate why identifying the special-function class of a differential equation is useful, but also why a special-function representation does not by itself solve the global spectral problem. Connection conditions, normalizability, and boundary conditions remain essential. \\
The present soft-core Dirac problem sits naturally at the intersection of these developments. Its reduced equation has several finite singular structures and an irregular point at infinity, while the original potential itself has algebraic branch points. One could therefore attempt a direct Heun-type classification. However, such a local classification would not by itself encode the full-line parity condition together with the global decay requirement. Instead, we use the branch structure to motivate a uniformizing coordinate and combine it with the exact \(Z_2\) symmetry of the Dirac Hamiltonian.\\
There is also a useful conceptual connection with the quantum Rabi model. Braak showed that its discrete symmetry can be used to organize the spectral problem into symmetry sectors and analytic spectral functions whose zeros encode eigenvalues \cite{Braak2011}. Subsequent work clarified continued-fraction formulations and their relation to analyticity \cite{BraakContinued2013}, introduced generalized $G$-functions with improved numerical control \cite{BraakGeneralized2013}, and extended the spectral-determinant viewpoint to the two-photon Rabi model, where the symmetry structure and pole properties of the $G$-function play an important role \cite{BraakTwoPhoton2023}. These studies are useful precedents for treating a spectral function as a global object rather than merely as a local special-function representation. Zhang\cite{ZhangAnalytic2014,ZhangHidden2016,ZhangKPhoton2017} independently developed complementary algebraic and analytic treatments of the Rabi family, including two-photon, two-mode, driven, and $k$-photon generalizations. His functional Bethe-ansatz treatment of the ordinary Rabi model and its two-photon and two-mode extensions is especially close in spirit to the QES benchmark used below \cite{ZhangRabi2013,ZhangAnalytic2014}. Our objective is not to identify the present problem with the Rabi model or to claim that its spectral function is Braak's $G$-function. Rather, we ask whether the same broad strategy---discrete symmetry followed by a global analyticity or connection condition---can be applied to a relativistic problem whose analytic structure is generated by physical soft-core Coulomb dynamics.\\
The specific Hamiltonian, its parity symmetry, and the uniformizing coordinate are introduced in the next section.
\section{Dirac Equation, Symmetry, and Uniformization}
We begin with the stationary Dirac equation in one spatial dimension. In the standard formulation of relativistic quantum mechanics, the general equation for a two-component spinor $\Psi=(F,G)^T$ can be written as \cite{GreinerRQM}
\begin{equation}
H\Psi(x)=E\Psi(x),
\qquad
H=\alpha p+\beta_D m+V(x)I,\qquad p=-i\frac{d}{dx},
\qquad
\alpha^2=\beta_D^2=I,
\quad
\{\alpha,\beta_D\}=0.
\label{eq:generalDirac}
\end{equation}
We choose the representation $\alpha=\sigma_x$ and $\beta_D=\sigma_z$. For the time-like vector potential considered below, this choice gives the coupled first-order component equations. We now specialize to the attractive soft-core Coulomb interaction $V(x)=-Z/\sqrt{x^2+\beta^2}$, with $Z>0$ and $\beta>0$, and set $m=\mu>0$ and $\hbar=c=1$. Here $V(x)I$ is a time-like vector (electrostatic) coupling, rather than a Lorentz-scalar coupling, which would instead enter through the mass term as $\beta_D S(x)$. With these choices, the general Dirac equation above becomes
\begin{equation}
F'(x)=i\left(E+\mu+\frac{Z}{\sqrt{x^2+\beta^2}}\right)G(x),
\qquad
G'(x)=i\left(E-\mu+\frac{Z}{\sqrt{x^2+\beta^2}}\right)F(x).
\label{eq:diraccomponents}
\end{equation}
This equation is the component form of the one-dimensional Dirac equation for the present soft-core Coulomb potential \cite{GreinerRQM}. For a bound state, \( -\mu<E<\mu \) and \(\kappa=\sqrt{\mu^2-E^2}>0\). Because the interaction is even, $V(-x)=V(x)$, the Hamiltonian has the exact discrete symmetry $\Pi=\sigma_zP_x$, with $\Pi^2=1$ and $[H,\Pi]=0$. This is the appropriate one-dimensional Dirac parity for the chosen representation: spatial reflection changes $p=-i\,d/dx$ to $-p$, while conjugation by $\sigma_z$ changes $\sigma_x$ to $-\sigma_x$. The problem therefore separates into two invariant parity sectors. If $\Pi\Psi=p\Psi$, $p=\pm1$, then
\begin{equation}
F(-x)=pF(x),\qquad G(-x)=-pG(x),
\label{eq:paritycomponents}
\end{equation}
so the two spinor components necessarily have opposite spatial parity. This use of even/odd spinor components is consistent with the general treatment of one-dimensional Dirac problems with even attractive potentials \cite{HallZorinNodal2014}.
Introduce the uniformizing coordinate
\begin{equation}
t=\sqrt{x^2+\beta^2},\qquad x=\sqrt{t^2-\beta^2}.
\end{equation}
The algebraic branch points \(x=\pm i\beta\) are mapped to \(t=0\), while the physical origin is mapped to \(t=\beta\).
For \(p=+1\), write \(F=f_+(t)\) and \(G=xg_+(t)\). Then
\begin{align}
f_+'(t)&=i\left[(E+\mu)t+Z\right]g_+(t),\label{eq:plus1}\\
(t^2-\beta^2)g_+'(t)+tg_+(t)&=i\left[(E-\mu)t+Z\right]f_+(t).\label{eq:plus2}
\end{align}
At \(t=\beta\), regularity gives \(g_+(\beta)=i(E-\mu+Z/\beta)f_+(\beta)\). Eliminating \(g_+\) and defining \(a=E+\mu\), \(b=E-\mu\), one obtains
\begin{equation}
(t^2-\beta^2)(at+Z)f_+''+(Zt+a\beta^2)f_+'+(bt+Z)(at+Z)^2f_+=0.
\label{eq:secondorder}
\end{equation}
Besides \(t=\pm\beta\), this equation has the energy-dependent finite singularity \(t_*=-Z/(E+\mu)\).
For \(p=-1\), writing \(F=xf_-(t)\) and \(G=g_-(t)\) gives
\begin{align}
(t^2-\beta^2)f_-'(t)+tf_-(t)&=i\left[(E+\mu)t+Z\right]g_-(t),\\
g_-'(t)&=i\left[(E-\mu)t+Z\right]f_-(t),
\end{align}
with the corresponding regularity condition at \(t=\beta\). Thus the exact parity constraint is imposed at the physical origin, while the uniformizing coordinate removes the square-root branch of the potential. The two operations have distinct roles: parity follows from the $Z_2$ symmetry of the original full-line Hamiltonian, whereas $t=\sqrt{x^2+\beta^2}$ is an analytic uniformization of the soft-core branch structure.
\section{Global Spectral Connection and Spectral Determinant}
The main purpose of the uniformization introduced above is not merely to rewrite the differential equation. It provides a convenient analytic domain in which the physical-origin condition and the large-distance bound-state condition can be treated as two ends of a single global connection problem. This is the point at which the present construction is inspired by the spectral-function strategy used for the quantum Rabi model and its extensions \cite{Braak2011,BraakContinued2013,BraakGeneralized2013,PengBraak2014,BraakTwoPhoton2023}. The analogy is structural: a discrete symmetry first selects independent sectors, and a sector-dependent global analytic condition then determines the spectrum. The spectral function introduced below is not Braak's $G$-function and is not obtained by transforming the Rabi model.
\subsection{Asymptotic decaying solution}
For $t\rightarrow\infty$, the soft-core interaction has the Coulombic expansion
\begin{equation}
-\frac{Z}{\sqrt{x^2+\beta^2}}
=-\frac{Z}{t}+O(t^{-3}).
\end{equation}
Thus the large-$t$ behavior is controlled by the same exponential decay associated with the mass gap as in the corresponding Coulomb tail, but it also contains an energy-dependent algebraic factor. Defining
\begin{equation}
\kappa=\sqrt{\mu^2-E^2}>0,
\qquad
\eta=\frac{EZ}{\kappa},
\end{equation}
the decaying solution can be written in the asymptotic form
\begin{equation}
\Psi_{\rm dec}(t;E)
\sim e^{-\kappa t}t^\eta
\left[
\bm u_0(E)+\frac{\bm u_1(E)}{t}
+\frac{\bm u_2(E)}{t^2}+\cdots
\right],
\label{eq:asymptotic}
\end{equation}
where the leading spinor may be normalized as
\begin{equation}
\bm u_0(E)=
\begin{pmatrix}
1\\[1mm]
i\kappa/(E+\mu)
\end{pmatrix}.
\label{eq:u0}
\end{equation}
The coefficients $\bm u_j(E)$ are obtained recursively by substituting the asymptotic series into the first-order Dirac system. Their explicit higher-order expressions are not required for defining the spectral function. The second independent asymptotic branch has the form $e^{+\kappa t}$ times an algebraic series and is therefore excluded from the bound-state subspace. It is useful to emphasize that the algebraic factor $t^\eta$ is essential. Omitting it would amount to treating the potential as short ranged and would give an incorrect large-distance connection problem. In particular, this factor is also the origin of the exponent appearing in the finite Bethe-ansatz construction discussed in Sec.~4.
\subsection{Parity-selected solution at the physical origin}
For each parity $p=\pm1$, the exact $Z_2$ symmetry fixes the relative parity of the two spinor components. A convenient normalized local solution is therefore generated at the physical point $t=\beta$. For $p=+1$ we use $F(0)=1$ and $G(0)=0$, equivalently
\begin{equation}
f_+(\beta)=1,
\qquad
g_+(\beta)=i\left(E-\mu+\frac{Z}{\beta}\right),
\label{eq:originplus}
\end{equation}
while for $p=-1$ one may use $F(0)=0$ and a nonzero normalization of $G(0)$. The overall normalization is irrelevant for the location of spectral zeros. Let $\Psi_p^{(\beta)}(t;E)$ denote the solution obtained by continuing the corresponding parity-selected local solution from $t=\beta$ toward the asymptotic region. For an arbitrary energy in the gap, this solution is a linear combination of the two independent asymptotic branches,
\begin{equation}
\Psi_p^{(\beta)}(t;E)
=
A_p(E)\Psi_{\rm dec}(t;E)
+
B_p(E)\Psi_{\rm grow}(t;E).
\label{eq:connection}
\end{equation}
The coefficients $A_p$ and $B_p$ depend on the chosen normalization of the local and asymptotic solutions, but the condition $B_p(E)=0$ does not depend on those nonzero normalizations.
\subsection{Connection function and Wronskian determinant}
The central object is the coefficient of the unwanted growing branch. In direct analogy with the role played by parity-resolved spectral functions in Braak's treatment of the quantum Rabi model, we define
\begin{equation}
\mathcal G_p(E)=B_p(E).
\label{eq:Gdef}
\end{equation}
A normalizable bound state exists precisely when the parity-selected origin solution contains no growing component,
\begin{equation}
\mathcal G_p(E)=0.
\label{eq:spectralcondition}
\end{equation}
Thus $\mathcal G_p(E)$ is a scalar, parity-resolved spectral function: its zeros determine the regular bound-state energies in the corresponding sector. This condition is global: neither the local equation near $t=\beta$ nor the asymptotic expansion alone determines it. It requires matching the two pieces of information across the full uniformized domain. The connection function can be written explicitly without extracting the coefficient $B_p(E)$ from the asymptotic expansion. For two spinors $\Psi_1=(F_1,G_1)^T$ and $\Psi_2=(F_2,G_2)^T$, define the two-component Wronskian
\begin{equation}
W[\Psi_1,\Psi_2]
=F_1G_2-G_1F_2.
\label{eq:wronskian}
\end{equation}
We then define the parity-resolved spectral determinant by
\begin{equation}
D_p(E)
=W\!\left[\Psi_p^{(\beta)}(E),\Psi_{\rm dec}(E)\right]_{t=t_0},
\label{eq:determinant}
\end{equation}
where $t_0$ is any point in the common domain of the two continued solutions. Because the coefficient matrix of the first-order Dirac system is trace-free, the Wronskian of two solutions is independent of the matching point. Consequently, the value of $D_p(E)$ is independent of $t_0$ apart from the harmless normalization chosen for the two solutions. To see directly why this determinant selects the spectrum, insert Eq.~\eqref{eq:connection} into Eq.~\eqref{eq:determinant}. Since $W[\Psi_{\rm dec},\Psi_{\rm dec}]=0$, one obtains
\begin{equation}
D_p(E)
=B_p(E)\,W[\Psi_{\rm grow},\Psi_{\rm dec}].
\label{eq:detconnection}
\end{equation}
With the asymptotic normalizations in Eqs.~\eqref{eq:asymptotic} and \eqref{eq:u0}, the leading spinors of the two branches give
\begin{equation}
W[\Psi_{\rm grow},\Psi_{\rm dec}]
=\frac{2i\kappa}{E+\mu},
\label{eq:asymptoticwronskian}
\end{equation}
and therefore the parity-resolved spectral function has the explicit representation
\begin{equation}
\boxed{%
\mathcal G_p(E)
=\frac{E+\mu}{2i\kappa}
W\!\left[\Psi_p^{(\beta)}(E),\Psi_{\rm dec}(E)\right]
=\frac{E+\mu}{2i\kappa}D_p(E).}
\label{eq:Gexplicit}
\end{equation}
Consequently,
\begin{equation}
\mathcal G_p(E)=0
\quad\Longleftrightarrow\quad
D_p(E)=0,
\label{eq:detcondition}
\end{equation}
provided the two asymptotic branches remain linearly independent. The overall prefactor in Eq.~\eqref{eq:Gexplicit} is a normalization convention and does not affect the spectral zeros. The terminology ``spectral determinant'' is used here in the connection sense: $D_p(E)$ is a determinant constructed from two independent global solutions whose vanishing expresses linear dependence between the physical-origin solution and the decaying solution. It should not be confused with a regularized functional determinant of an infinite-dimensional operator. This distinction is useful because the present construction is intended to provide a concrete spectral function that can be evaluated numerically and, where possible, analytically.
\subsection{Relation to the Braak construction}
The connection with Braak's approach is structural rather than an identification of the two spectral problems. In the quantum Rabi model, the conserved $Z_2$ symmetry separates the Hilbert space into invariant parity sectors, and analyticity of the corresponding Bargmann-space solutions leads to parity-resolved spectral functions whose zeros determine the eigenvalues \cite{Braak2011}. Continued-fraction and generalized-$G$-function formulations develop this analytic spectral condition in alternative forms \cite{BraakContinued2013,BraakGeneralized2013}. Related extensions to multi-qubit and two-photon Rabi models further illustrate how parity or higher discrete symmetries, analyticity, exceptional solutions, and spectral-collapse structure enter generalized spectral constructions \cite{PengBraak2014,FelicettiBraak2015,BraakTwoPhoton2023}.
For the present Dirac soft-core problem, the analogous sequence is
\begin{equation}
\begin{aligned}
Z_2 &\longrightarrow \text{parity sector}
\longrightarrow \text{uniformized analytic domain} \\
&\longrightarrow \text{origin--infinity connection}
\longrightarrow \mathcal G_p(E)=0.
\end{aligned}
\label{eq:braakchain}
\end{equation}
Here, however, the first step is the exact real-space symmetry $\Pi=\sigma_zP_x$, and the spectral condition is obtained by matching a parity-selected physical solution at the origin to the decaying solution at infinity. The nontrivial analytic structure is generated by the soft-core branch points and the first-order Dirac system, not by a Bargmann-space representation. Consequently, $\mathcal G_p(E)$ is not Braak's $G$-function, and the present construction is not claimed to be equivalent to Braak's solution. We refer to $\mathcal G_p(E)$ as a Braak-inspired parity-resolved spectral function, while $D_p(E)$ denotes its Wronskian normalization. The useful analogy is the organization of the spectrum by a discrete symmetry together with a sector-dependent global spectral condition.
The distinction also clarifies the role of the Bethe ansatz. A finite Bethe polynomial produces a special QES solution for which the global connection coefficient can vanish analytically. Such a state is therefore a zero of $D_p(E)$ when it satisfies the physical parity condition, but a zero of the global determinant does not in general require finite polynomial termination. The Bethe sector is thus a restricted algebraic sector embedded within the broader global spectral problem.
\subsection{Numerical evaluation of the determinant}
For numerical purposes, one may construct $\Psi_p^{(\beta)}$ by integrating outward from $t=\beta$ and construct $\Psi_{\rm dec}$ by integrating inward from a sufficiently large $t=L$. The determinant is then evaluated at an intermediate matching point $t_m$,
\begin{equation}
D_p(E;t_m)=
F_p^{(\beta)}(t_m)G_{\rm dec}(t_m)
-G_p^{(\beta)}(t_m)F_{\rm dec}(t_m).
\label{eq:numericaldet}
\end{equation}
For an exact calculation, $D_p(E;t_m)$ is independent of $t_m$ up to numerical error. This provides an internal stability check: changing $t_m$ should leave the zero of $D_p(E)$ unchanged while the absolute value of the determinant may vary with normalization. For the numerical implementation used below, we evaluate the determinant itself rather than using the finite-$L$ residual as the primary spectral function. For the $p=+1$ sector, the origin solution can be normalized as $\boldsymbol{\Psi}_+^{(\beta)}(0)=(1,0)^T$. Hence, at the origin,
\begin{equation}
D_+(E)=W[\boldsymbol{\Psi}_+^{(\beta)},\boldsymbol{\Psi}_{\rm dec}]_{x=0}=G_{\rm dec}(0;E),
\label{eq:Dplusorigin}
\end{equation}
up to the chosen normalization of the decaying solution. In practice it is numerically more stable to use the normalized determinant
\begin{equation}
\widehat D_+(E)=\frac{D_+(E)}{F_{\rm dec}(0;E)}
=\frac{G_{\rm dec}(0;E)}{F_{\rm dec}(0;E)},
\label{eq:normalizeddet}
\end{equation}
whose zeros are exactly the zeros of $D_+$ whenever $F_{\rm dec}(0;E)\neq0$. We obtain the decaying solution by integrating inward from a large $x=L$, with its leading asymptotic ratio imposed at the starting point. The zero of $\widehat D_+(E)$ is then located directly. Thus the numerical energies reported below are zeros of the spectral determinant, not zeros of the auxiliary finite-distance residual.
\section{Bethe-Ansatz Benchmark and Numerical Validation}
The global determinant can be tested against a finite Bethe-ansatz sector. For the $p=+1$ sector, Eq.~\eqref{eq:secondorder} is used directly, following the functional Bethe-ansatz strategy developed by Zhang for second-order differential equations and subsequently applied to the quantum Rabi model and its two-photon, two-mode, and $k$-photon generalizations \cite{ZhangPolynomial2012,ZhangRabi2013,ZhangAnalytic2014,ZhangHidden2016,ZhangKPhoton2017}. Related relativistic soft-core Coulomb constructions provide an additional QES precedent \cite{AgboolaZhang2012,AgboolaZhang2013}. We write
\begin{equation}
 f_n(t)=e^{-\kappa t}t^{\rho_n}(t+\beta)^{1/2}P_n(t),
\qquad
P_n(t)=\prod_{j=1}^{n}(t-t_j),
\label{eq:BAansatz}
\end{equation}
where
\begin{equation}
\kappa=\sqrt{\mu^2-E^2}.
\end{equation}
The large-$t$ balance is especially useful because it gives a general parametric form for the Bethe-ansatz energy. Since $P_n(t)\sim t^n$ and the Coulomb tail produces the algebraic factor $t^{EZ/\kappa}$, one obtains
\begin{equation}
\rho_n+n+\frac12=\frac{EZ}{\kappa}.
\label{eq:rho_general}
\end{equation}
Defining
\begin{equation}
N_n=n+\frac12+\rho_n,
\end{equation}
Eq.~\eqref{eq:rho_general} becomes
\begin{equation}
\frac{EZ}{\sqrt{\mu^2-E^2}}=N_n.
\end{equation}
Thus the positive-energy branch can be written parametrically as
\begin{equation}
E_n=\frac{\mu N_n}{\sqrt{Z^2+N_n^2}},
\qquad
N_n=n+\frac12+\rho_n.
\label{eq:generalBAenergy}
\end{equation}
Changing the overall sign in this parametrization gives the formal negative-energy counterpart, but this does not imply an exact $E\to -E$ symmetry of the present vector-coupled Dirac Hamiltonian. Physical negative-energy bound states, if present, must be found by imposing the full-line boundary condition independently in the interval $-\mu<E<0$. Equation~\eqref{eq:generalBAenergy} is not, by itself, a universal spectrum: $\rho_n$, $\beta$, and the Bethe roots must still satisfy the finite coefficient and boundary conditions. It is the general energy parametrization supplied by the large-$t$ condition. A particularly transparent QES subclass is obtained when
\begin{equation}
\rho_n=0.
\end{equation}
Then
\begin{equation}
E_n=\frac{\mu(n+1/2)}{\sqrt{Z^2+(n+1/2)^2}}
=\frac{\mu(2n+1)}{\sqrt{(2n+1)^2+4Z^2}}.
\label{eq:rhozeroenergy}
\end{equation}
This relation must be accompanied by the remaining termination conditions and therefore should not be interpreted as a spectrum valid for arbitrary $\beta$.
Substitution of Eq.~\eqref{eq:BAansatz} into Eq.~\eqref{eq:secondorder}, followed by evaluation at a zero $t_i$ of $P_n$, gives the Bethe equations
\begin{equation}
2\sum_{\substack{j=1\\j\neq i}}^n
\frac{1}{t_i-t_j}
-2\kappa
+\frac{2\rho_n}{t_i}
+\frac{1}{t_i+\beta}
+
\frac{Zt_i+(E+\mu)\beta^2}
{(t_i^2-\beta^2)[(E+\mu)t_i+Z]}
=0 .
\label{eq:Bethe}
\end{equation}
These equations are necessary but not sufficient. The remaining coefficients of the substituted differential equation must vanish, and the resulting polynomial must also satisfy the physical origin condition. The latter point is essential: a polynomial solution of the second-order equation is not automatically a bound state of the full-line Dirac problem.

For $n=0$, $P_0=1$. Choosing the simple QES branch $\rho_0=0$ gives
\begin{equation}
E_0=\frac{\mu/2}{\sqrt{Z^2+1/4}}.
\end{equation}
Choosing units such that $\mu=1$ and taking $Z=1/2$ gives
\begin{equation}
E_0=\frac{1}{\sqrt{2}},
\qquad
\kappa_0=\frac{1}{\sqrt{2}},
\qquad
\beta=1,
\end{equation}
with
\begin{equation}
f_0(t)=e^{-t/\sqrt{2}}\sqrt{t+1}.
\label{eq:n0exact}
\end{equation}
The corresponding first-order Dirac component is reconstructed from the original Dirac equation. This solution satisfies the $p=+1$ origin condition and therefore constitutes a complete analytic bound-state benchmark rather than merely a local solution of the second-order equation. The QES examples below are benchmark constructions for different choices of the Hamiltonian parameters. In particular, the $n=1$ and $n=2$ parameter sets are not additional states of the fixed Hamiltonian $(\mu,Z,\beta)=(1,1/2,1)$ used in Sec.~5; they correspond to separate admissible soft-core Hamiltonians. For $n=1$, take $P_1(t)=t+c$. The termination equations admit an algebraic QES branch with
\begin{equation}
E_1=\frac45,
\qquad
\kappa_1=\frac35,
\qquad
Z=\frac98,
\qquad
\rho_1=0,
\end{equation}
and $c$ satisfying
\begin{equation}
4608c^3-14400c^2-9000c+38125=0,
\label{eq:n1cubic}
\end{equation}
The cubic has three real roots. Two of them give admissible positive values of the soft-core parameter. The third root, $c=1.94894856$, gives $\beta=-2.59423597$ and is therefore excluded by $\beta>0$. Direct evaluation of the full-line parity condition confirms that the two positive-$\beta$ branches belong to the $p=+1$ sector. Thus the $n=1$ Bethe-ansatz termination conditions provide two distinct admissible parameter sets with the same energy $E_1=0.80000000$, as summarized in Table~\ref{tab:n1BAbranches}.
\begin{table}[ht]
\centering
\caption{Admissible $n=1$ Bethe-ansatz branches for the present Dirac soft-core problem. Both branches have $\mu=1$, $Z=9/8$, $\rho_1=0$, and $E_1=0.80000000$, but correspond to different positive soft-core parameters.}
\label{tab:n1BAbranches}
\begin{tabular}{c c c c}
\toprule
$c$ & $\beta$ & $E_1$ & parity $p$\\
\midrule
$-1.55462754$ & $1.09723540$ & $0.80000000$ & $+1$\\
$ 2.73067899$ & $4.62200057$ & $0.80000000$ & $+1$\\
\bottomrule
\end{tabular}
\end{table}

For $n=2$, writing $P_2(t)=t^2+c_1t+c_0$, the same procedure gives algebraic termination conditions. One branch is specified by
\begin{equation}
E_2=\frac45,
\qquad
\kappa_2=\frac35,
\qquad
Z=\frac{15}{8},
\qquad
\rho_2=0,
\end{equation}
and the coefficient $c_1$ satisfies
\begin{align}
&41472c_1^5-432000c_1^4-2880000c_1^3
+18825000c_1^2 \nonumber\\
&\qquad +97343750c_1+104765625=0.
\label{eq:n2quintic}
\end{align}
For example, the real root
\begin{equation}
c_1=-2.15864686
\end{equation}
leads to
\begin{equation}
\beta=4.28950686,
\qquad
c_0=-12.84372731.
\end{equation}
As for $n=1$, this is an algebraic QES solution of the reduced equation; its status as a physical full-line state has to be decided only after the parity boundary condition is imposed. We therefore use the $n=0$ solution as the strict full-line benchmark in the present numerical comparison and retain the $n=1,2$ constructions as explicit analytical QES benchmarks. The global connection construction provides the independent test. For a genuine QES bound state, the analytic energy must satisfy
\begin{equation}
D_+(E_n^{\rm BA})=0,
\label{eq:BAcheck}
\end{equation}
where $D_+$ is the Wronskian determinant defined in Sec.~3. Conversely, a zero of $D_+$ need not belong to the finite Bethe sector. This is the precise sense in which the Bethe ansatz is used here as a benchmark rather than as the definition of the full spectrum. For the direct numerical calculation, we integrate the original first-order Dirac equations after writing $G=ig$,
\begin{equation}
F'=-\left(E+\mu+\frac{Z}{\sqrt{x^2+\beta^2}}\right)g,
\qquad
 g'=\left(E-\mu+\frac{Z}{\sqrt{x^2+\beta^2}}\right)F.
\label{eq:numericalfirstorder}
\end{equation}
For $p=+1$ the origin solution is normalized by $F(0)=1$, $g(0)=0$. The decaying solution is integrated inward from a large $x=L$ using
\begin{equation}
\frac{g(L)}{F(L)}=\frac{\kappa}{E+\mu}
\end{equation}
as its leading asymptotic condition. Equivalently, one may integrate the ratio $r(x)=g(x)/F(x)$, which obeys
\begin{equation}
r'=E-\mu+\frac{Z}{\sqrt{x^2+\beta^2}}
+\left(E+\mu+\frac{Z}{\sqrt{x^2+\beta^2}}\right)r^2.
\label{eq:riccati}
\end{equation}
At the origin the normalized determinant is simply
\begin{equation}
\widehat D_+(E)=r(0).
\label{eq:detriccati}
\end{equation}
Thus the numerical bound-state energy is obtained from the zero of the spectral determinant, $\widehat D_+(E)=0$.
For the exact $n=0$ benchmark,
\begin{equation}
\mu=1,\qquad Z=\frac12,\qquad \beta=1,
\end{equation}
and the Bethe-ansatz prediction is
\begin{equation}
E_0^{\rm BA}=\frac{1}{\sqrt2}=0.70710678.
\end{equation}
For a numerically stable evaluation of both parity sectors, it is useful to write the decaying solution in polar form,
\begin{equation}
F=R\cos\theta,\qquad g=R\sin\theta.
\end{equation}
The angle then satisfies
\begin{equation}
\theta'=\left(E-\mu+\frac{Z}{\sqrt{x^2+\beta^2}}\right)\cos^2\theta
+\left(E+\mu+\frac{Z}{\sqrt{x^2+\beta^2}}\right)\sin^2\theta,
\label{eq:angleequation}
\end{equation}
with the decaying asymptotic condition
\begin{equation}
\theta(L)=\arctan\!\left(\frac{\kappa}{E+\mu}\right).
\end{equation}
At the physical origin, the two parity conditions become
\begin{equation}
p=+1:\quad \sin\theta(0)=0,
\qquad
p=-1:\quad \cos\theta(0)=0.
\end{equation}
Thus the normalized Wronskian determinants can be represented numerically as
\begin{equation}
\widehat D_+(E)=\sin\theta(0;E),
\qquad
\widehat D_-(E)=-\cos\theta(0;E),
\label{eq:angledeterminants}
\end{equation}
which have exactly the same zeros as the corresponding Wronskian connection determinants, provided the same decaying solution and normalization are used.

For the $n=0$ QES parameter set, the determinant zero in the $p=+1$ sector converges to $0.70710678$, reproducing the analytic Bethe-ansatz value. The same fixed Hamiltonian also possesses additional determinant zeros that are not generated by the displayed finite-polynomial QES construction. Using $L=100$, the first five positive-energy determinant zeros are
\begin{equation}
\begin{aligned}
p=+1:&\quad
E=0.70710678,\quad 0.95074219,\quad 0.98111275,\\
p=-1:&\quad
E=0.90297038,\quad 0.97160987.
\end{aligned}
\label{eq:extendedspectrum}
\end{equation}
The first value is the analytic $n=0$ Bethe-ansatz benchmark. The remaining values are obtained from the global spectral determinant for the same fixed Hamiltonian and do not arise from the $\rho_n=0$ finite-polynomial QES sector used for that benchmark. They therefore provide a direct numerical illustration of the fact that the spectral determinant is not exhausted by the Bethe-ansatz construction. The determinant zero was also checked against the outer integration point. For the $n=0$ state, changing $L$ from $30$ to $200$ leaves the energy unchanged at the displayed eight-decimal precision. The small differences from the analytic value are numerical integration and root-finding errors rather than physical shifts.
\begin{table}[ht]
\centering
\caption{Spectral-determinant benchmark for the $n=0$ QES bound state. The last column gives the absolute difference between the determinant zero and the analytic Bethe-ansatz value.}
\label{tab:determinant_difference}
\begin{tabular}{c c c}
\toprule
$L$ & $E_0^{D}$ & $\left|E_0^{D}-E_0^{\rm BA}\right|$\\
\midrule
30  & 0.70710678 & $5.03375119\times10^{-13}$\\
50  & 0.70710678 & $7.38520356\times10^{-13}$\\
100 & 0.70710678 & $6.75348666\times10^{-13}$\\
200 & 0.70710678 & $1.53288493\times10^{-12}$\\
\bottomrule
\end{tabular}
\end{table}
\begin{table}[ht]
\centering
\caption{Positive-energy bound-state zeros of the global spectral determinant for the fixed parameter set $\mu=1$, $Z=1/2$, and $\beta=1$. The state $E=0.70710678$ is the analytic $n=0$ Bethe-ansatz benchmark; the other listed states are obtained directly from the parity-resolved determinant.}
\label{tab:beyondBA}
\begin{tabular}{c c c}
\toprule
Parity $p$ & $E$ & QES benchmark in the present sector\\
\midrule
$+1$ & 0.70710678 & $n=0$\\
$-1$ & 0.90297038 & ---\\
$+1$ & 0.95074219 & ---\\
$-1$ & 0.97160987 & ---\\
$+1$ & 0.98111275 & ---\\
\bottomrule
\end{tabular}
\end{table}
These additional roots should be interpreted as a numerical demonstration of the broader domain of the global determinant, not as a claim that the displayed list exhausts the entire spectrum. In particular, states close to the continuum threshold require larger $L$ and tighter numerical control. The Bethe-ansatz construction remains an exact QES benchmark, including the two admissible $n=1$ branches in Table~\ref{tab:n1BAbranches}, whereas the determinant is defined independently by the global origin--infinity connection. The numerical spectrum displayed in Table~\ref{tab:beyondBA} was searched only in the positive-energy gap $0<E<\mu$; no claim of spectral completeness or of an exact $E\to-E$ symmetry is made.\\
The relation to the quantum Rabi model is structural. In Braak's construction, the $Z_2$ symmetry leads to parity-resolved analytic spectral functions, with continued-fraction, generalized-$G$, and later generalized spectral-determinant formulations providing complementary realizations of the same global spectral philosophy \cite{Braak2011,BraakContinued2013,BraakGeneralized2013,BraakTwoPhoton2023}. Here, the corresponding sequence is
\begin{equation}
Z_2\longrightarrow\text{parity sector}\longrightarrow\text{uniformization}
\longrightarrow\text{global connection}\longrightarrow D_\pm(E)=0.
\end{equation}
The Bethe-ansatz benchmark provides a complementary local analytic check, while the additional determinant zeros in Table~\ref{tab:beyondBA} demonstrate explicitly that the global spectral construction is not restricted to the finite QES sector.
\section{Conclusions}
We have formulated the one-dimensional Dirac soft-core Coulomb problem as a parity-resolved global connection problem. The attractive potential retains the physical long-range Coulomb behavior while replacing the origin singularity by the finite soft-core scale \(\beta\). Its algebraic branch points are uniformized by \(t=\sqrt{x^2+\beta^2}\), and the exact \(Z_2\) symmetry generated by \(\Pi=\sigma_zP_x\) provides the natural decomposition into two parity sectors. For each parity, a solution selected at the physical origin can be connected to the decaying and growing solutions at infinity. The coefficient of the growing branch defines a spectral connection function, equivalently represented by a Wronskian determinant. Its zeros provide the bound-state condition. An independent Bethe-ansatz construction was then derived directly from the same Dirac equation. The resulting quasi-exact solutions provide analytic benchmarks for the global spectral condition. The important distinction is that the Bethe ansatz describes only a restricted QES sector, whereas the spectral determinant is defined independently and is intended to characterize the complete spectrum. \\
The resulting framework is inspired by the structural role of $Z_2$ symmetry and sector-resolved spectral functions in Braak's solution of the quantum Rabi model \cite{Braak2011}. The present problem has a different analytic structure and a different mechanism for imposing the spectral condition: the determinant is constructed from a real-space origin--infinity connection rather than Bargmann-space analyticity. It should therefore not be identified with Braak's $G$-function or regarded as an equivalent formulation of the Rabi solution. The combination of a physical soft-core Coulomb interaction, algebraic uniformization, exact spinor parity, and a global connection determinant provides a concrete framework for relativistic spectral problems with nontrivial analytic structure.
\\section*{Acknowledgments}
The author acknowledges the use of OpenAI's ChatGPT as an artificial-intelligence-assisted tool during the preparation of this manuscript, including for language editing, organization of the presentation and checking. The author takes full responsibility for the scientific content, calculations, interpretations, and conclusions presented in the paper.

\end{document}